\documentclass[ pra,twocolumn,superscriptaddress]{revtex4-1}
\usepackage[colorlinks=true, urlcolor=blue, pdfborder={0 0 0}]{hyperref}

\usepackage{graphicx}
\usepackage{amsmath}
\usepackage{mathrsfs}
\usepackage{amsfonts}
\usepackage{color}
\usepackage{bm} 
\usepackage{subfigure}
\usepackage{epsfig} 
\usepackage{epstopdf}
\usepackage{float} 
\usepackage{tikz}
\usetikzlibrary{shapes.arrows}
\usepackage[percent]{overpic}
\usepackage{comment}
\usepackage{pstricks}
\usepackage{amssymb}

\usepackage[utf8]{inputenc}

\def\beq{\begin{equation}}
\def\eeq{\end{equation}}
\def\beqn{\begin{eqnarray}}
\def\eeqn{\end{eqnarray}}

\begin{document}

\title{\bf Probing Quantum States with Momentum Boosts}
\author{Tarek A. Elsayed}
\thanks{Author to whom correspondence should be addressed;\\ \href{mailto:telsayed@zewailcity.edu.eg}{telsayed@zewailcity.edu.eg}, \href{mailto:tarek.ahmed.elsayed@gmail.com}{tarek.ahmed.elsayed@gmail.com}}
\address{Zewail City of Science and Technology, 6th of October City, Giza 12578, Egypt}
\address{Department of Physics and Astronomy, Aarhus University, 8000 Aarhus C, Denmark}
\address{Theoretische Chemie,  Physikalisch-Chemisches Institut, Im Neuenheimer Feld 229, D-69120 Heidelberg, Germany}

\author{ Alexej I. Streltsov}
\address{Theoretische Chemie,  Physikalisch-Chemisches Institut, Im Neuenheimer Feld 229, D-69120 Heidelberg, Germany}
\address{Institut für Physik, Universität Kassel, Heinrich-Plett-Str. 40, 34132 Kassel, Germany}

\date{\today}
\begin{abstract}
We present a technique to diagnose the condensate fraction in a one-dimensional optical lattice of weakly interacting 
bosons based on the dynamics of the trapped atoms under the influence of a momentum kick. It is shown using the Multi-Configuration Time Dependent Hartree method for Bosons (MCTDHB) that the two extreme cases of the superfluid and Mott insulator states exhibit different behaviors when the lattice is briefly tilted. The current induced by the momentum boost caused by the tilt which depends directly on the amount of phase coherence between the lattice sites is linearly proportional to the condensate fraction. The atom-atom interactions only change the slope of the linear relationship. We discuss the applications of this scheme in magnetic field gradiometery.

\end{abstract}
\maketitle



\section{Introduction}

The advances in the physics of trapped ultracold atoms has led to a fast growing spectrum of applications such as quantum computers and simulators \cite{ladd2010,debnath2016}, 
high precision measurement and quantum metrology  \cite{abend2016, peters2001,min2015} and testing the fundamental theories of physics \cite{stedman1997}. Ultracold atomic clouds immersed in an optical lattice is a basic paradigm  of  quantum systems used in these applications \cite{bloch2005,jaksch1998}.  Many of the  recent experiments on ultracold atomic clouds in optical lattices involve a time-dependent manipulation of the underlying trapping potential (e.g., tilting) or the interaction between the atoms (e.g., quenching).  Developing a good understanding of these experiments requires  a detailed knowledge of the properties of the involved quantum states at both the qualitative and the quantitative level. 

The most important classes of the quantum states of ultracold atoms in optical lattices are the superfluid (SF) state and the Mott insulator (MI) state. The transition between these two limits at zero temperature defines a quantum phase transition \cite{greiner2002,jaksch1998}. Since the first experimental realization of Bose-Einstein Condensates (BECs) in optical lattices, one of the most important quantities used to characterize the superfluid-to-Mott-insulator transition has been the noncondensed fraction \cite{greiner2002}. This quantity is also relevant for studying the fragmentation in BECs with degenerate ground state \cite{mueller2006}. In the case of a few-well systems with a considerable number of atoms per site, the Mott-insulating
physics boils down to the fragmentation phenomenon. For the double-well system \cite{gati2006,gati2006_2},  the Mott-like state corresponds to a two-fold fragmented state. 

Fragmentation properties can be strictly defined in terms of the reduced one-particle density \cite{penrose1956,nozieres1982,mueller2006} :
$\rho^{(1)}(x,x')= \left \langle \hat{\Psi}^\dag(x')\hat\Psi(x) \right \rangle$ where $\hat\Psi$ is the field operator. 
For a pure state, we have 
$\rho^{(1)}(x_1,x'_1;t)\!\!= \!\!N\int \Psi^\ast(x'_1,x_2,\ldots,x_N;t) \Psi(x_1,x_2,\ldots,x_N;t) dx_2\cdots dx_N$, where $N$ is the number of atoms and $\Psi$ is the many-body wavefunction. This matrix can also be defined in terms of  its eigenvalues (natural occupations) $n_k$ and eigenvectors (natural orbitals) $\phi_k$ as
\beq
\label{RDM}
\rho^{(1)}(x,x';t)=\sum^M_{k=1} n_k(t) \phi_k(x,t)\phi_k^*(x’,t).
\eeq
For a thermal state, the occupancies $n_k$ are temperature-dependent. The system is called {\it condensed} if only one natural orbital is macroscopically occupied $n_1\!\!\approx\! \!N$ \cite{penrose1956} and {\it fragmented} if several eigenvalues have macroscopic occupations \cite{nozieres1982}. The condensed fraction $n$ refers to the relative occupation of the most occupied natural orbital $n=\frac{n_1}{N}$, while the noncondensed fraction is $(1-n)$.

Unlike the SF state where the quantum states exhibits a well defined phase at every lattice site, the MI state exhibits no global phase coherence  but rather a well-defined number of atoms at each lattice site. Therefore, the ballistic expansion method  has been the main  tool to distinguish between the two phases and  quantify the noncondensed fraction in trapped ultracold systems \cite{bloch2005}. In this method, the visibility of the interference fringes formed during the ballistic expansion after the atoms are released to free space allows one to quantify the (non)condensed fraction of the original quantum state \cite{simon2008}.   In principle, a single experimental measurement (a single shot) allows to quantify the noncondensed fraction in an optical lattice. 
 For few-well systems, the proper measurement of the visibility in a fragmented system requires a multiple repetition of the full experimental  sequence with a careful statistical analysis \cite{gati2006,gati2006_2}. In general, this method suffers from several effects such as  a finite time-of-flight and an inhomogenous trapping potential \cite{gerbier2007}.
 
Some alternative methods have been proposed to probe the condensate fraction and the superfluid transition such as Bragg spectroscopy \cite{stenger1999,inada2008} or the quantum microscope which measures single site-resolved atom number fluctuations \cite{bakr2010}. It was recently also reported that measuring the density-density correlations after free expansion is a good  signature of the fragmentation in single-site traps \cite{kang2014}. Other methods (such as  \cite{streltsova2014,kronke2015}) are more involved and include additional manipulations on the clouds e.g., phase-imprinting with more advanced statistical collection and repetition and post-processing of the experimental shots.

The aim of this paper is to provide a less demanding alternative for measuring the condensate fraction in systems of weakly interacting ultracold atoms based on the dynamical behavior of the single-particle density. We show that the tunneling behavior of the atoms in the optical lattice under the influence of a momentum kick has a strong dependence on $n$.  We  analyze the dynamics of two- and many-site one-dimensional optical lattice systems described by the Hamiltonian $H=\sum_i \left( \frac{P_i^2}{2m}+V(x_i)  \right)+\sum_{i<j}W(x_i-x_j) $, where $m$ is the mass of the atom, $P_i$ is the momentum of atom $i$, $V(x)$ is the lattice potential and $W(x_i-x_j) $ is the interatomic interaction potential. We use a periodic lattice of the form $V(x)=V_0 \cos(\frac{\pi x}{a})$ where $V_0$ is the depth of the lattice and $a$ is the lattice constant while  we take $W(x_i-x_j)$ as a delta-function potential (i.e., hard-core interaction)  $W(x_i-x_j)=W_0\times\delta(x_i-x_j)$. For all the numerical simulations in this paper, we use the Multi-Configurational Time-Dependent Hartree method for Bosons (MCTDHB) \cite{streltsov2007,alon2008,lode2012}  which is available within the MCTDHB-Laboratory package \cite{mctdh-b}. We  give a prescription for quantifying the condensate fraction in terms of the atomic oscillations after the lattice is briefly tilted. In order to simulate quantum states with different condensate fractions, we use the ground state of an interacting system  with different values of  $W_0$ as initial states. Whenever we refer to interatomic interactions, we mean for the actual boosting experiment unless otherwise stated.


This paper is structured as follows. In section \ref{Sec2}, we consider the simplest two-well model system and introduce the key concept behind our method. Section \ref{Sec3} describes how to diagnose the condensate fraction in an unknown stationary quantum state for a multi-well optical lattice and discusses the effect of the atom-atom interactions. In section \ref{Sec4}, we show a side application of our method to the field of quantum metrology. Section \ref{Sec5} concludes and summarizes our study.

\section{Tunneling Dynamics in a Double-Well Potential}\label{Sec2}

The double-well system is a fundamental building block for studying  correlations and tunneling dynamics of quantum many-body systems that has been realized with large controllability for bosonic \cite {folling2007} and fermionic atoms \cite{murmann2015}. As simple as it is, this system captures the physics of basic solid state models such as the Hubbard model and can exhibit rich dynamics (see for example \cite{smerzi1997}). In this section, we illustrate the method proposed in this paper for a double-well potential and show  the effect of the coherence between the two wells on the tunneling dynamics through the barriers.

\begin{figure*}[t] \setlength{\unitlength}{0.1cm}
\begin{picture}(160 , 85 ) 
{

\put(17, 0)  {\includegraphics[ width=2cm,height=1.5cm]{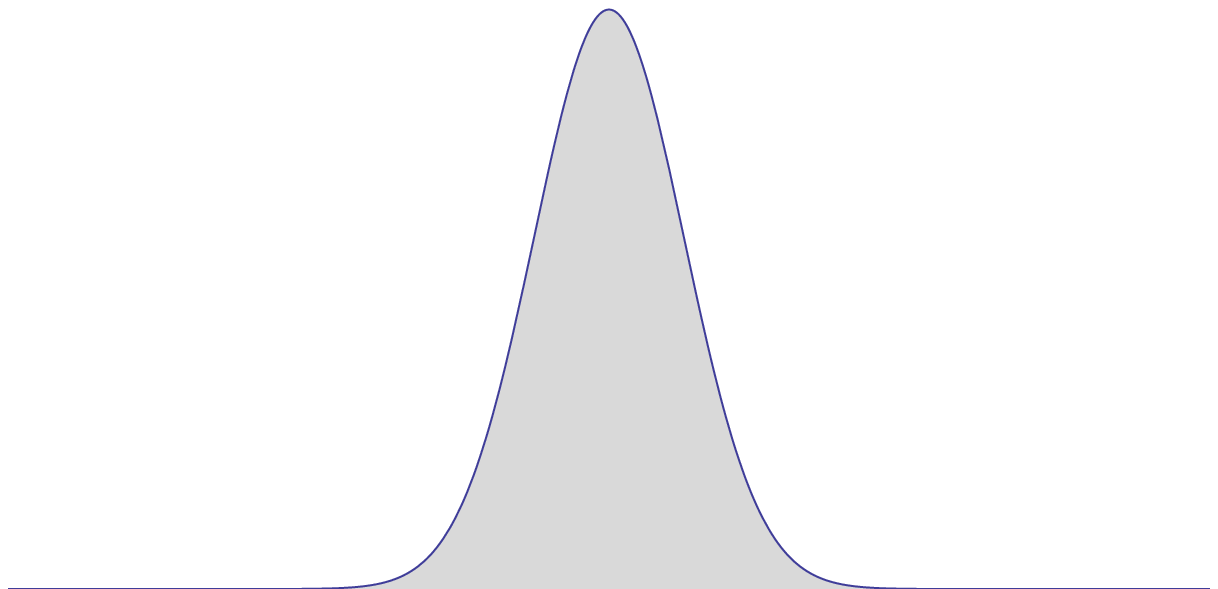}}

\put(-3, 17)  {\includegraphics[ width=2cm,height=1.0cm]{packet.eps}}

\put(17, 17)  {\includegraphics[ width=2cm,height=1.0cm]{packet.eps}}

\put(-3, 34)  {\includegraphics[ width=2cm,height=1.5cm]{packet.eps}}

\put(17, 51)  {\includegraphics[ width=2cm,height=1.0cm]{packet.eps}}

\put(-3, 51)  {\includegraphics[ width=2cm,height=1.0cm]{packet.eps}}

\put(17, 68)  {\includegraphics[ width=2cm,height=1.5cm]{packet.eps}}

\put(25, 60)  { $e^{i 0}$}
\put(4, 60)  {$e^{-i \pi/2}$}

\put(25, 26)  {$e^{i 0}$}
\put(4, 26)  {$e^{i \pi/2}$}
\put(16, 80)  {(a)}

\put(50, 43)  {\includegraphics[ width=5.5cm,height=3.7cm]{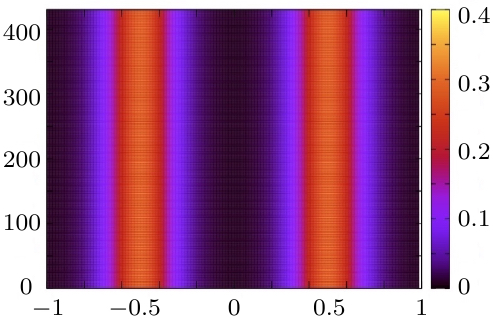}}

\put(50, 1)  {\includegraphics[ width=5.5cm,height=3.7cm]{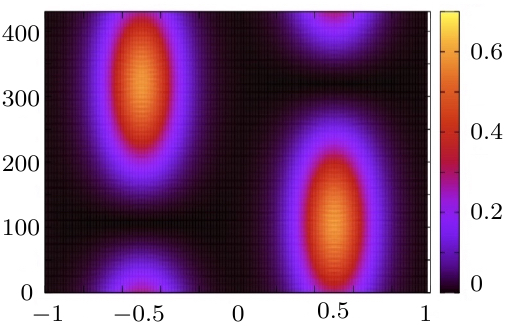}}

\scriptsize
\put(73, -1) {$ { x/a}$ }
\put(73, 41) {$ { x/a}$ }
\put(47, 16) {\rotatebox{90} {Time (s)}}
\put(47, 55) {\rotatebox{90} {Time (s)}}

\put(131, 73)  {\epsfig{file=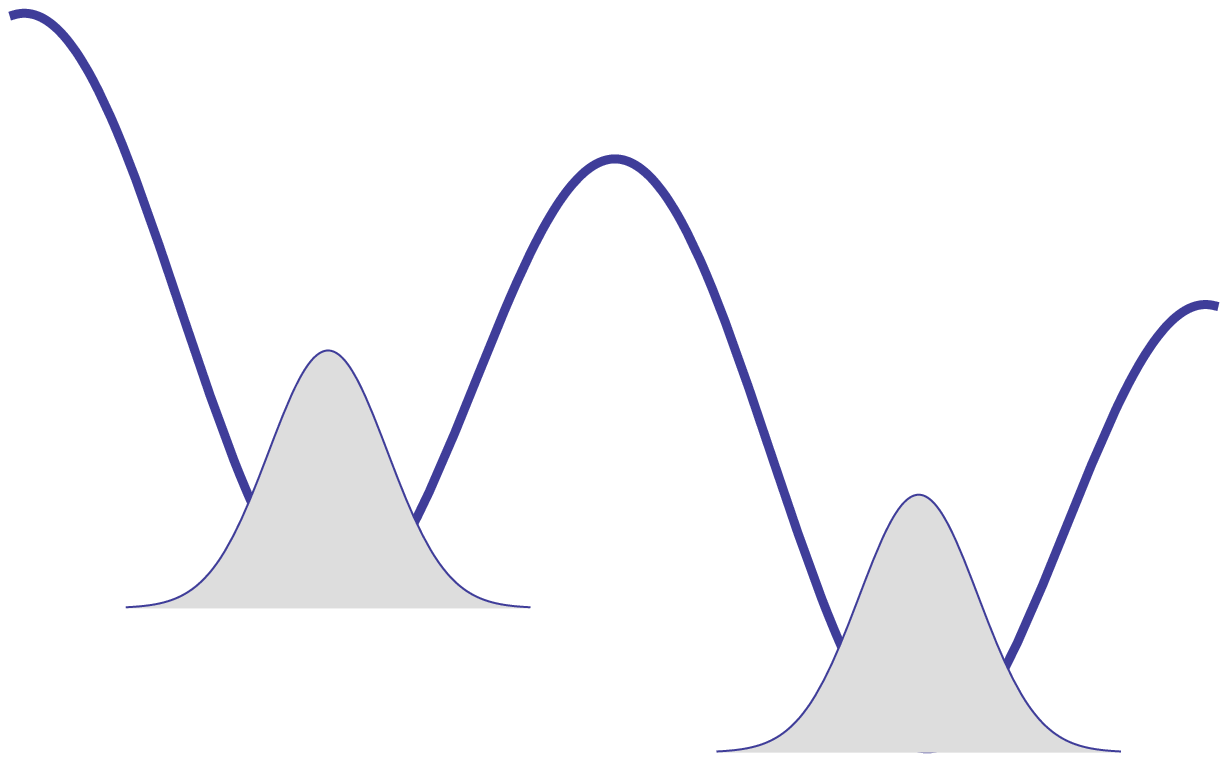,scale=0.1} }

\put(145, 73)  {\includegraphics[ scale=0.1]{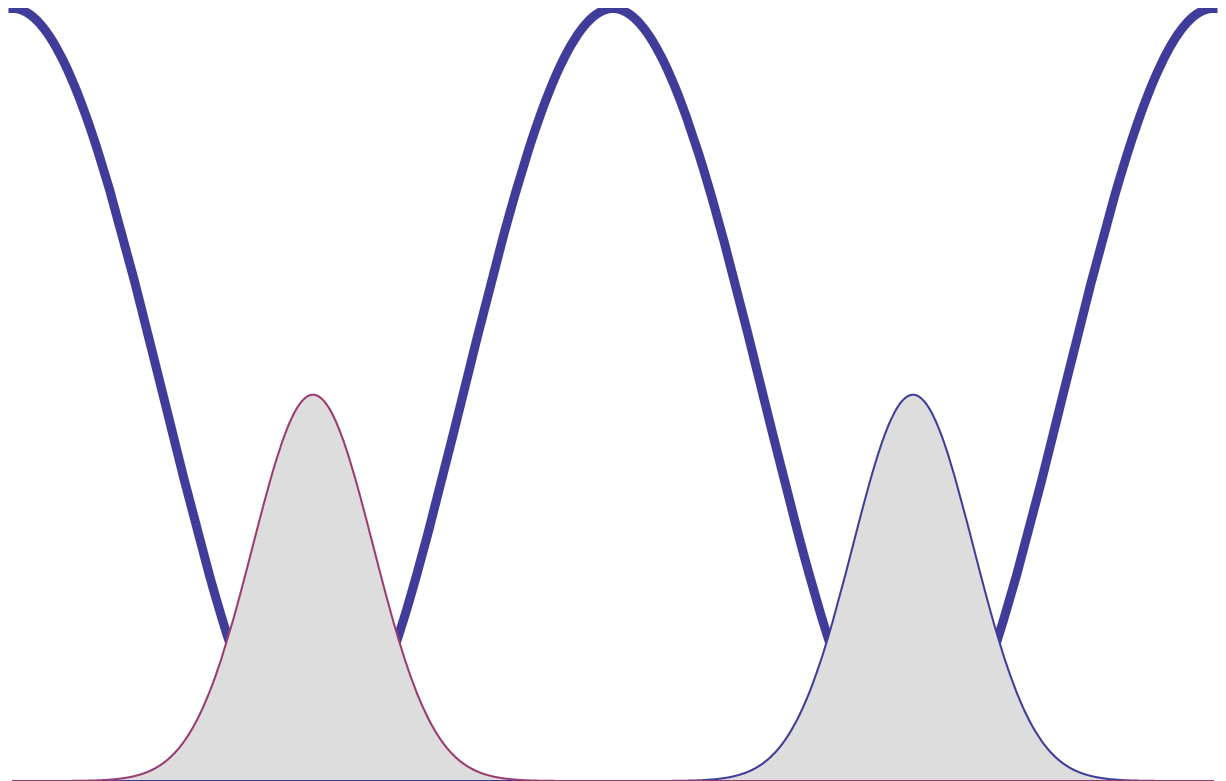}}

\put(125, 0)  {\includegraphics[scale=0.3]{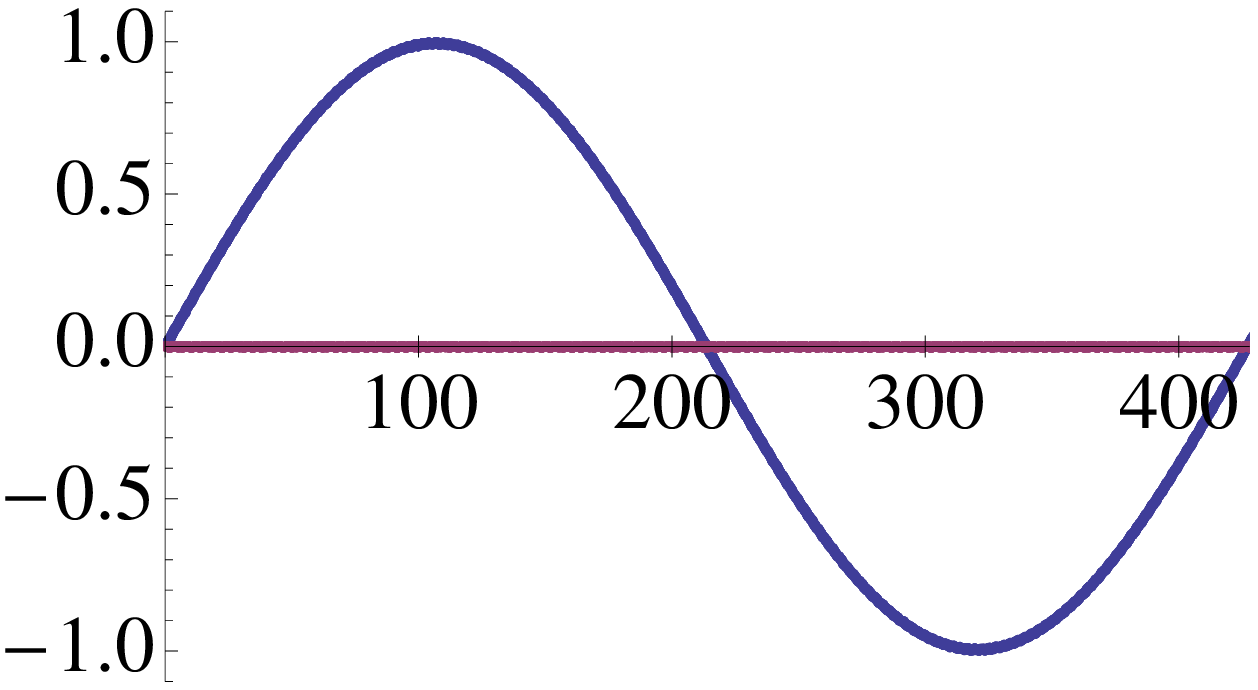}}

\put(125, 23)  {\includegraphics[scale=0.3]{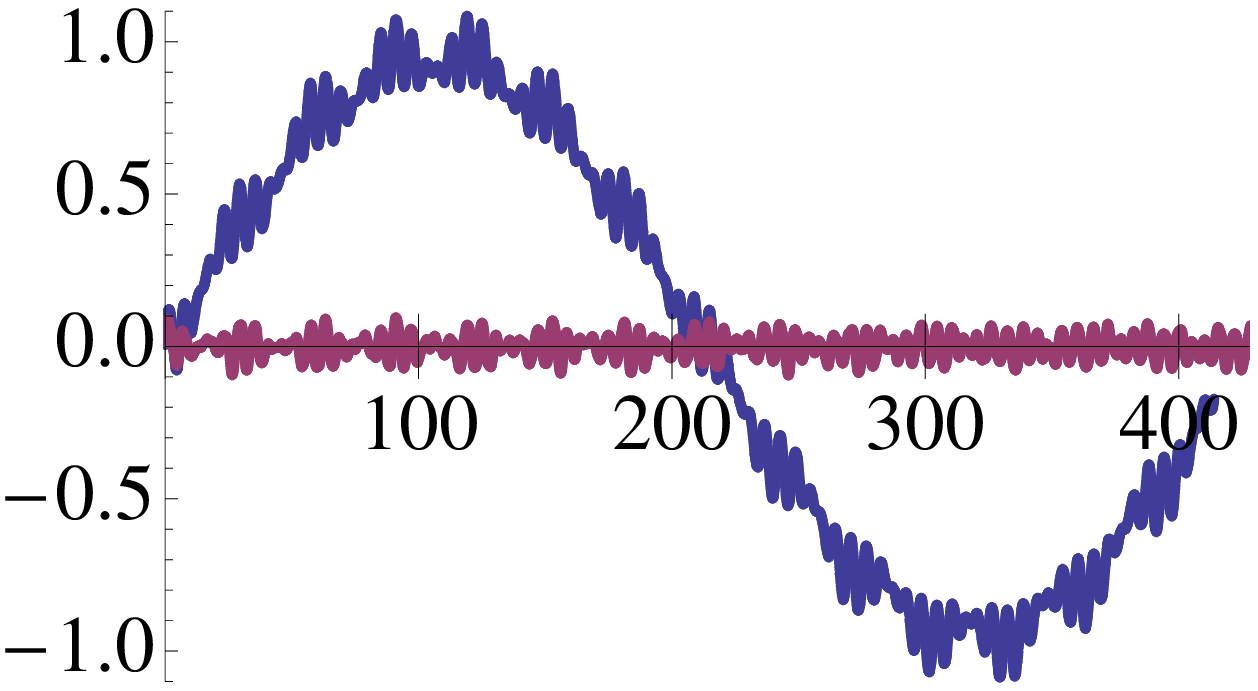}}

\put(125, 46)  {\includegraphics[ scale=0.3]{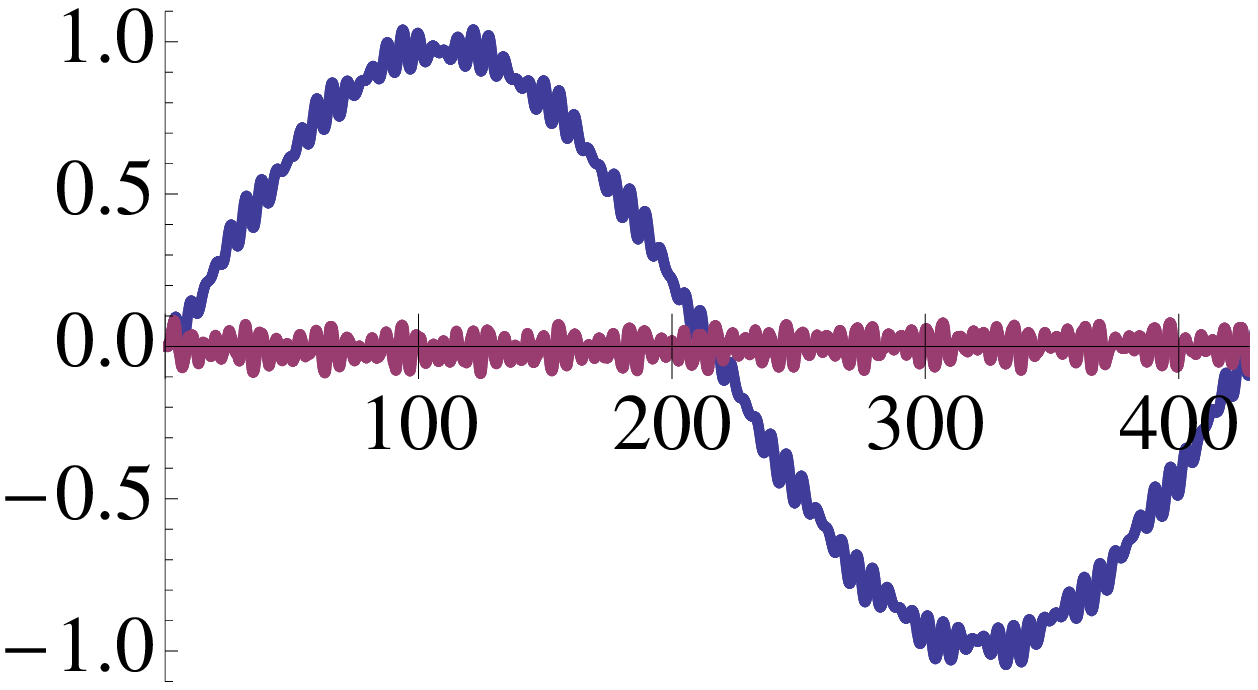}}

\put(137, 4.5) {Time (s)}
\put(137, 27) {Time (s)}
\put(137, 50) {Time (s)}

\put(118, 19.5) {$\langle x/a \rangle$}
\put(118, 41.5) {$\langle x/a \rangle$}
\put(118, 65) {$\langle x/a \rangle$}

\put(107, 23)  {(c)}
\put(107, 60)  {(b)}

\put(157, 62)  {(d)}
\put(157, 40)  {(e)}
\put(157, 16)  {(f)}

\put(142, 81)  {\Large $\curvearrowleft$}

\put(143, 72)  {\Large \rotatebox{180} {$\curvearrowleft$}}
\put(144, 84)  {I}
\put(144, 67)  {II}

}
\end{picture} 

\caption{ \label{fig:Fig1} Tunneling dynamics in a double well potential. (a) The evolution of a condensate localized in a single well during a complete Rabi cycle, indicating the phase difference between the two sites during the delocalized phases. (b,c) Minkovski space-time plots of the density of atoms in a double well in the  Mott-like and superfluid states respectively. The two systems are initially given a momentum boost by imprinting a phase difference of $\pi/2$ on the wavefunction in the two sites.  (d-f) The mean value of the position of the atoms $\langle \hat{x} \rangle$ for the superfluid state (blue) and the Mott state (red) when the momentum boost is generated by a brief tilt in the lattice (see upper panel) (d), imprinting a phase gradient $e^{ikx}$ (e), or directly imprinting a $\pi/2$ phase difference between the sites of the two wells in the initial state (f). The tilt and phase gradients are adjusted to lead to a  $\pi/2$ phase difference between the sites as well.   All distances are scaled by $a$, the distance between the two wells. }
\end{figure*}

 To understand the effect of coherence on tunneling, consider a double-well system with two non-interacting atoms. If this system is initialized  such that the two atoms are condensed in the ground state of one well (see Fig.~\ref{fig:Fig1}-a), the atoms will eventually tunnel through the barrier to the other well. During this process, the atoms will pass through an intermediate (superfluid) phase where they are equally distributed in a coherent superposition between the two wells. Therefore, if the system is initialized in this superfluid state with the proper phase relation between the two wells,  a current will be induced in the lattice (Fig.~\ref{fig:Fig1}-c).
 
  On the other hand, if this system is initialized in a Mott-like state, an incoherent superposition of two condensates, each localized in one well (i.e., each atom is not aware of the other atom), each atom will  tunnel through the barrier in a direction opposite to the other one and the net current will be zero (Fig.~\ref{fig:Fig1}-b). This can be understood mathematically by noticing that imprinting a phase difference between the two sites causes only a global phase change to the wavefunction of the MI state, and hence leads to no physical changes.

The contrast between these two extreme cases in terms of the current induced in the lattice is a measurable effect that can be used to distinguish between them. To induce this current, one needs to create the initial phase relation between the two wells.
The current induced  will be maximum when the initial phase difference between the two sites of the double-well is $\pi/2$ as in the transient  cases in Fig.~\ref{fig:Fig1}-a. We can induce this current in the numerical simulation by imprinting this phase difference manually to the initial state, or by imprinting a phase gradient $e^{ikx}$ such that $ka=\pi/2$ where $a$ is the distance between the two wells. Imprinting this phase gradient is equivalent to giving a momentum boost to the system. This can be physically realized  by briefly tilting the lattice at the beginning. Tilting the optical lattice potential means  adding 
a potential gradient $V_\text{tilt}=\gamma x$ to the optical lattice potential. 
This potential can be of the same type of the lattice potential, or a different type, i.e., a gravitational potential. The duration of the tilt should be short compared to the timescale of the atomic dynamics. In order for the tilt to achieve the same effect, 
the phase difference  between the two sites developed during the tilt interval $T$ 
should be of  the same magnitude, i.e., $\gamma a T/\hbar=\pi/2$.

 \normalsize

In Figs.~\ref{fig:Fig1}-d,~\ref{fig:Fig1}-e and ~\ref{fig:Fig1}-f, we show the atomic oscillations in double-wells for the superfluid and Mott states with noninteracting atoms using the three different boost mechanisms presented above. Namely, the phase differences between the two wells are created by either performing the tilt mechanism (d), imprinting a phase gradient (e) or imprinting a discrete phase difference (f). The last case is the theoretical case consistent with the intermediate figures in panel (a), while the first two ones are simulations of more practical situations. In this simulation we used a lattice potential of the form $V(x)=2 \cos^2(0.2\pi x)$, where $V_0$ can be expressed in terms of the recoil energy $E_R$ ($E_R=\frac{h^2}{2m\lambda^2}$, $\lambda$ is the wavelength of the laser used to construct the lattice and $\lambda=2a$) as $V_0=101 E_R$ (we use $\hbar=1$ and $m=1$ in all numerical simulations in this paper). This trapping potenial gives rise to a tunneling coefficient $J=h\times  11.7\times 10^{-3}\ \text{Hz}$, where $h$ is Planck's constant, corresponding to a Rabi cycle of $427$ s  (see, e.g.,  \cite{jaksch1998} for details of computing $J$).

We notice in Fig. \ref{fig:Fig1}-d, -e, and -f  the existence of fast fluctuating oscillations with small amplitude on top of $\langle x \rangle$ in (d) and (e) corresponding to fast oscillations within each well probably due to the coupling to higher bands  (see section 4).   Apart from these fluctuations,   the current indicated by the oscillations of the mean distance of the atoms $\langle \hat{x} \rangle$ with respect to the center of the double well potential is almost the same for the three studied superfluid states and zero for the Mott states.  This distinct behavior implies that the intermediate cases where only a fraction of the atoms is in the superfluid state yield intermediate values for the current induced, and hence this current can be used to probe the condensate fraction $n$.

We verify this observation in Fig.~\ref{fig:Fig2}.
 In this figure, we show the atomic oscillations in the same double well in Fig.~\ref{fig:Fig1}, for different  initial states having different condensate fractions $n$. These states are taken to be the ground states of Hamiltonians with different inter-particle interaction strengths. After the initial states are generated, the interaction is switched off, and the boost is implemented using the tilt mechanism. There is no net induced current in these states before applying the momentum boost.  These results were obtained by MCTDHB with 2 orbitals (MCTDHB(2)) but they are identical to MCTDHB(4). The waveforms of $\langle \hat{x} \rangle $ are smoothed out by a moving average technique for the sake of clarity. In Fig.~\ref{fig:Fig2}-b, we notice the remarkable linear dependence of the amplitude 
of the oscillations on the condensate fraction $n$.
Since this scheme requires the atomic dynamics to be primarily due to the momentum boosts, it is suitable only for diagnosing stationary quantum states.

\section{Diagnosis of the condensate fraction in an optical lattice} \label{Sec3}

The technique presented in the previous section for a double-well potential is directly generalizable to an optical lattice. To understand  the difference between the superfluid and  Mott states in terms of  the currents induced, consider the two cases in the second quantization picture. We shall suppose without loss of generality that the occupation is one atom per site. The superfluid wavefunction is expressed as 
\begin{equation}
 \Psi_\text{SF}=\left(\sum_{i=1}^N \hat{a}_i^\dag\right)^N |0\rangle
 \end{equation} 
while the Mott insulator wavefunction in the atomic limit (zero tunneling) is expressed as
\begin{equation}
 \Psi_\text{MI}=\prod_{i=1}^N \hat{a}_i^\dag |0\rangle,
 \label{MI}
 \end{equation} 
where $\hat{a}_i^\dag$ is the bosonic creation operator at site $i$.

The current operator is 
\begin{equation}
\hat{J}=\frac{\hbar}{2mi}\left( \hat{\Psi}^\dag(r)\nabla\hat{\Psi}(r)-\nabla\hat{\Psi}^\dag(r)\hat{\Psi}(r)\right)
\label{current}
 \end{equation} 
where  $\hat{\Psi}(r)$ is the field operator. 
If we take our basis to be the Wannier states $\{ w_n(r)\} $ which are localized at each lattice site, i.e., $w_n(r)=w(r-r_n) $, we can express the field operator as 
\begin{equation}
\hat{\Psi}(r)=\sum_{n}\hat{a}_n w_n(r).
 \end{equation} 
In the theoretical case where the phase of the wavefunction at each lattice site is given a different value in discrete steps of $\Delta \phi$, we can let this phase be encoded in the Wannier states, i.e., $\tilde{w}_n=e^{in\Delta \phi}w_n$. 
Substituting the field operator in \ref{current} in terms of he new Wannier states, we find that
\begin{widetext}
\begin{equation}
\langle \hat{J} \rangle =  \frac{\hbar}{2mi}\left( \sum_k \int \tilde{w}_k^*(r)\nabla \tilde{w}_k(r)dr\langle a^\dag_k a_k \rangle +\sum_{k\neq q} \int \tilde{w}_q^*(r)\nabla \tilde{w}_k(r)dr\langle a^\dag_q a_k \rangle -c.c\right)
 \end{equation} 
\end{widetext}
The first term on the right-hand side (RHS) does not contribute in the current due to the vanishing integral. The second term vanishes in the case of  a Mott insulator due to the absence of correlations between different sites, i.e.,  it can be verified using \ref{MI} that $\langle a^\dag_q a_k \rangle_\text{MI}=0$ for $k\neq q$. This is not the case, however, in the superfluid state. If we consider the contribution from the integral in the second term on the RHS to be predominantly due to neighboring sites, we find that in the superfluid state, $\langle \hat{J} \rangle$ is proportional to $\sin (\Delta \phi)$.

 \begin{figure}[] \setlength{\unitlength}{0.1cm} 
\begin{picture}(75 , 28 ) 
{
\put(-4, 0)  {\includegraphics[ scale = 0.3]{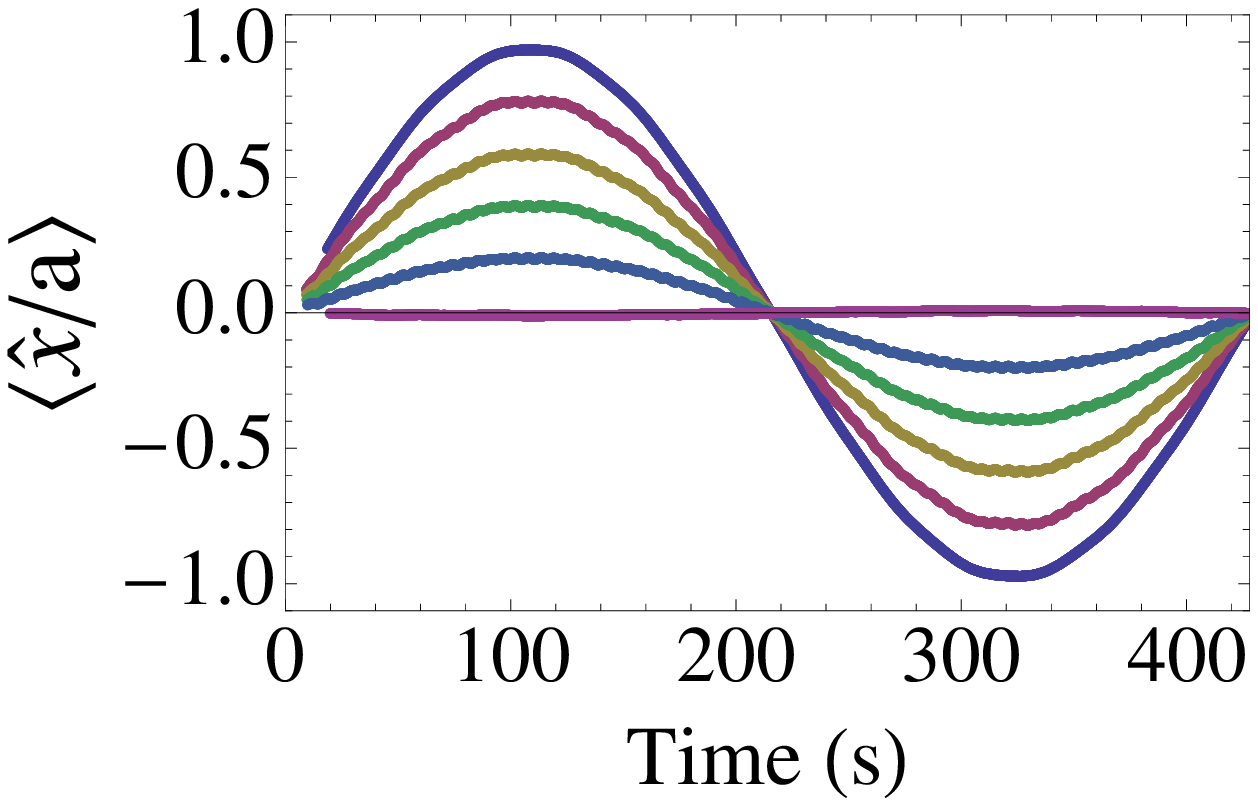}}
\put(37, 0)  {\includegraphics[ scale = 0.3]{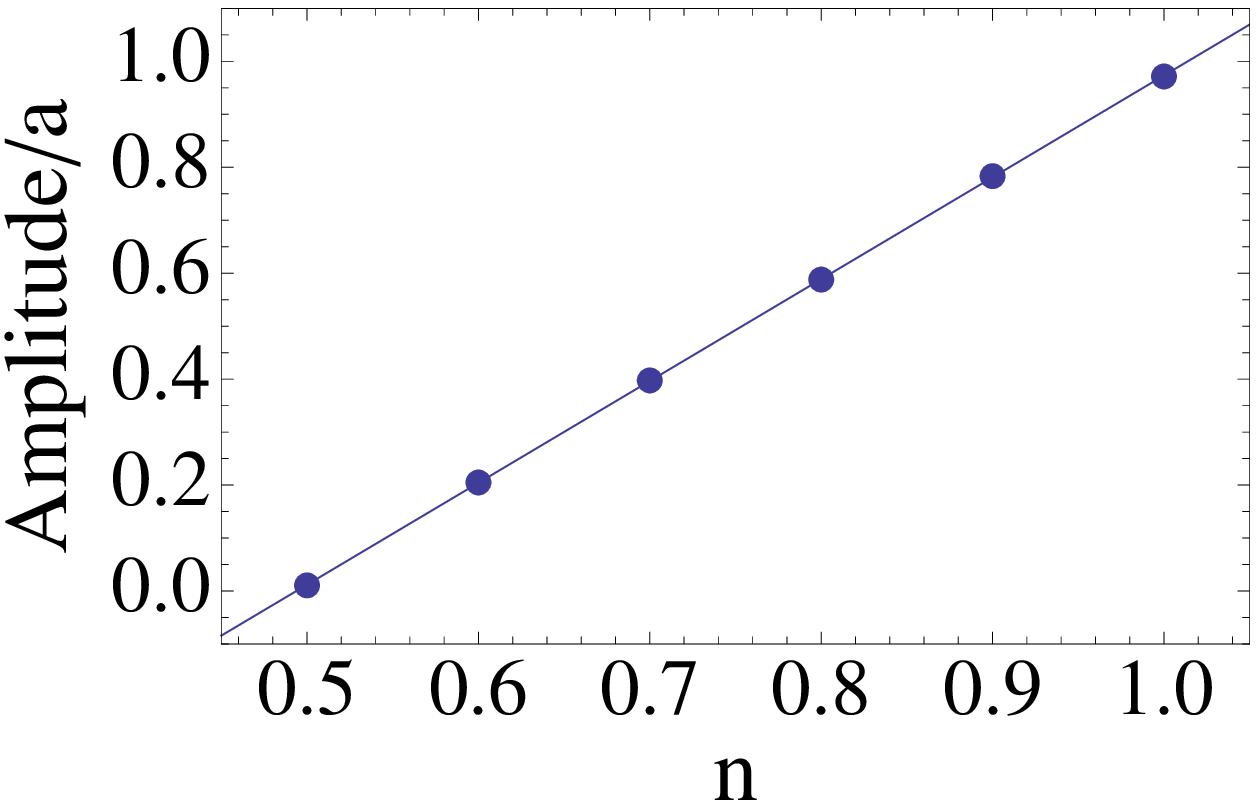}}

\put(15, 26) {\footnotesize (a)}
\put(54, 27) {\footnotesize(b)}

}
\end{picture} 

\caption{ \label{fig:Fig2}  (a) Matter wave oscillations in a double-well potential subjected to a brief tilting initially for different condensation fraction. (b) The  amplitude of the oscillations in (a) versus the  condensate fraction $n$. Distances are normalized with respect to $a$, the separation  between the two wells.}
\end{figure}

\begin{figure}[] \setlength{\unitlength}{0.1cm}
\begin{picture}(80 , 80 ) 
{
\put(-4, 64)  {\includegraphics[ scale = 0.3]{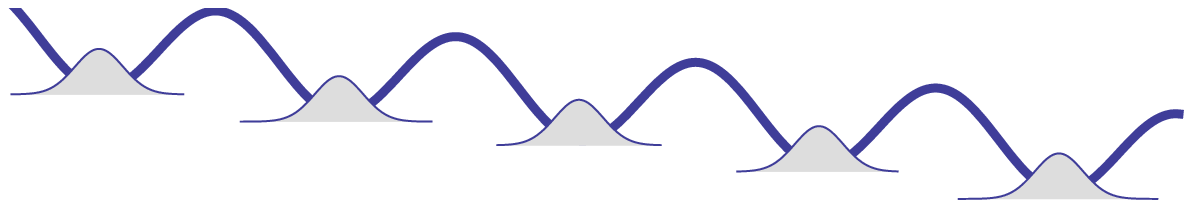}}
\put(37, 66)  {\includegraphics[ scale = 0.3]{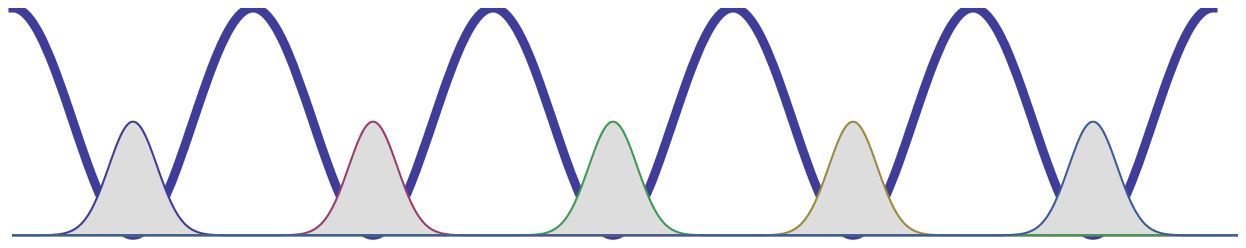}}

\put(-1, 30)  {\includegraphics[ scale = 0.22]{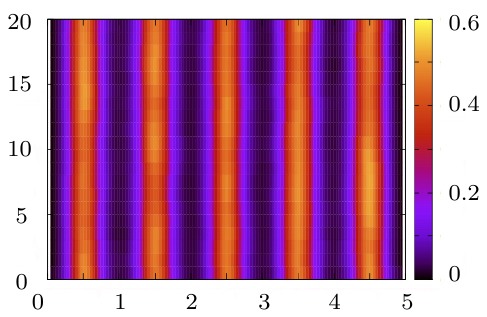}}

\put(-1, 2)  {\includegraphics[ scale = 0.17]{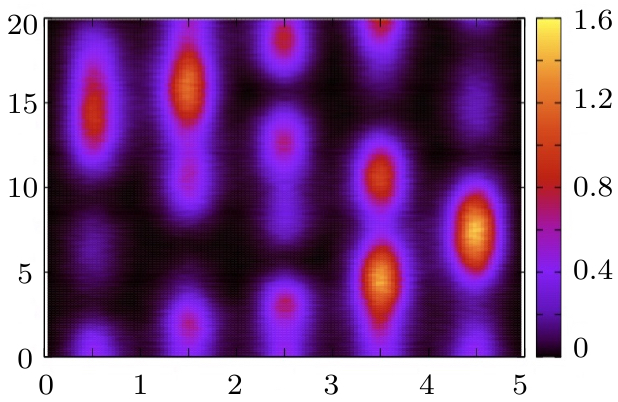}}

\put(40, 29)  {\includegraphics[ scale = 0.31]{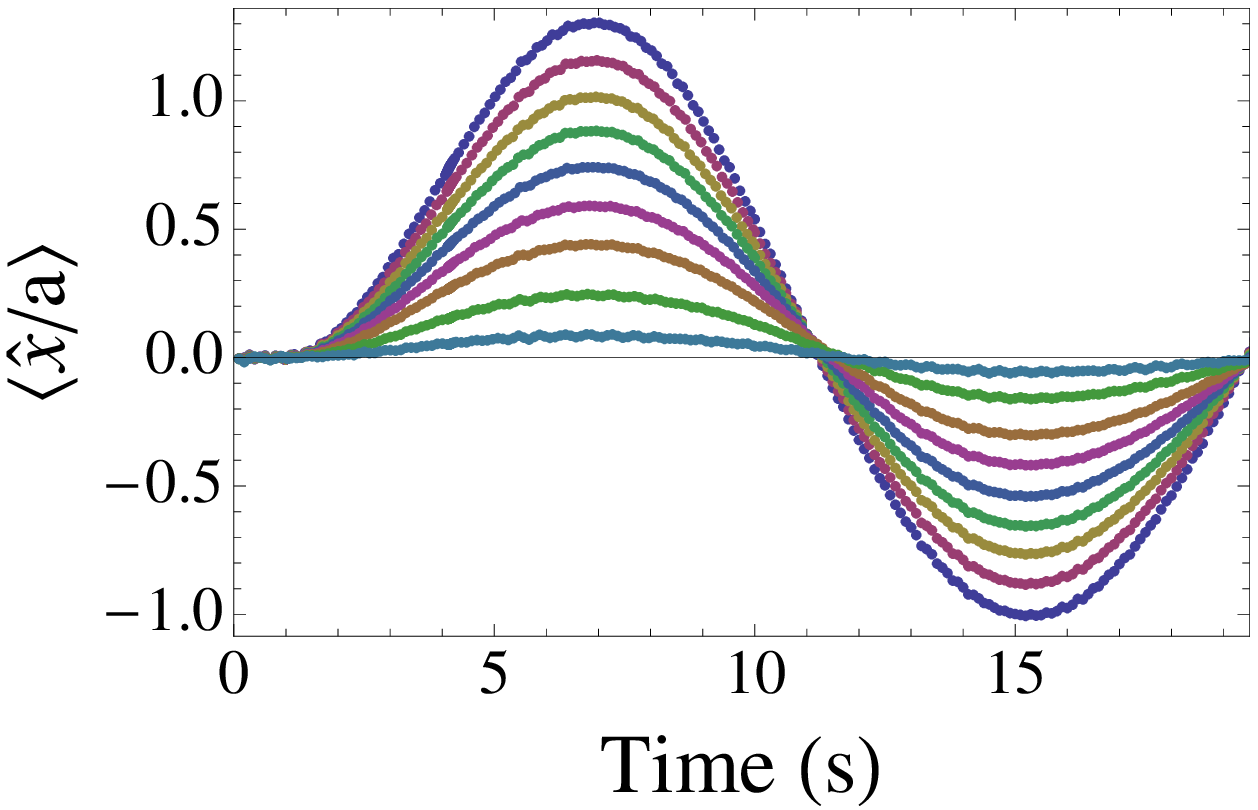}}
\put(40, 0)  {\includegraphics[ scale = 0.31]{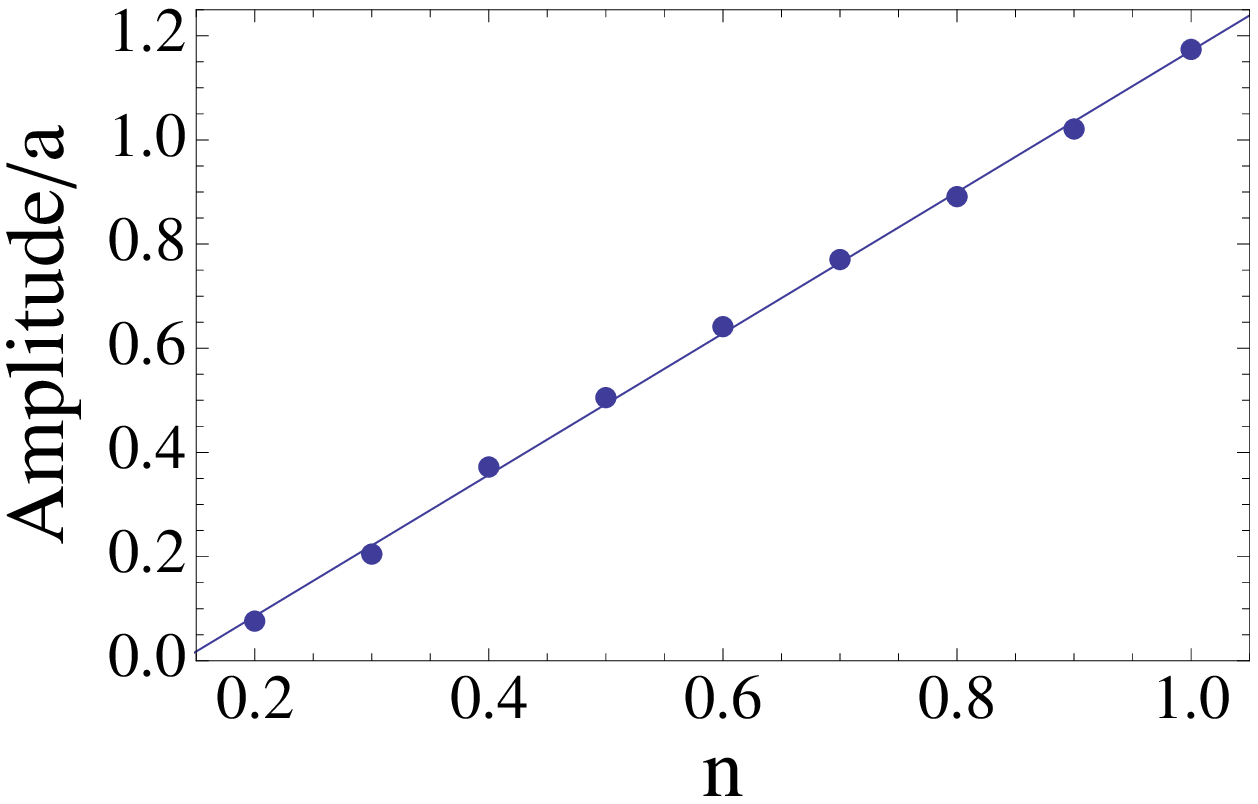}}

\put(14, 0) {$ {\scriptstyle x/a}$ }
\put(14, 28) {$ {\scriptstyle x/a}$ }
\put(-4, 12) {\rotatebox{90} {\scriptsize Time (s)}}
\put(-4, 39) {\rotatebox{90} { \scriptsize Time (s)}}

\footnotesize
\put(-2, 56) {\scalebox{.8}{(a)}}
\put(-2, 27.5) {\scalebox{.8} {(b)}}

\put(39, 56) {\scalebox{.8} {(c)}}
\put(39, 27.5) {\scalebox{.8}{(d)}}
\normalsize

\put(32, 74)  {\Large $\curvearrowleft$}
\put(34, 65)  {\Large \rotatebox{180} {$\curvearrowleft$}}
\put(34, 79)  {\small I}
\put(35, 59.5)  {\small II}

}
\end{picture} 

\caption{ \label{fig:Fig3} Diagnosing the condensate fraction in a 5-site optical lattice with non-interacting atoms. (a-b) Minkovski space-time plots of the density of  atoms after a momentum boost caused by a brief tilting of the lattice in Mott state and a superfluid state respectively (see the upper panel).  (c) The oscillation of the atoms for different condensation fractions. (d) The  amplitude of the oscillations in (c) versus the  condensate fraction $n$. Distances are normalized with respect to $a$, the separation  between the two wells.}
\end{figure}

In Fig.~\ref{fig:Fig3}, we illustrate the mean atomic displacement in a 5-site optical  lattice of the form $V(x)=25\cos^2(\pi x)$, corresponding to $V_0=5E_R$, without inter-atomic interaction. The lattice contains 5 atoms, and we used 5 orbitals in the simulation. The initial states are again taken to be the ground states of interacting Hamiltonians with different interaction strengths $W_0$ ranging from 0 to 25.  We briefly tilt the lattice potential in the beginning of the simulation to imprint a phase difference of $\pi/2$ to the wavefunction between each neighboring lattice sites. The closest state to a Mott insulator with maximum fragmentation we could achieve has 20\% occupation for each natural orbital. This state corresponds to the lowest point in Fig.~\ref{fig:Fig3}-d. On the other hand, the superfluid state with maximum occupation of a single natural orbital corresponds to the highest point in Fig.~\ref{fig:Fig3}-d with maximum amplitude of the current. 

The pronounced linear behavior in the dependence of the current on the fragmentation in Fig.~\ref{fig:Fig2} and Fig.~\ref{fig:Fig3} indicates that in order to diagnose the condensate fraction in an optical lattice in an unknown state, 
it suffices to identify the two points on this graph  corresponding to the extreme cases of superfluid and Mott states. Reliable simulations of larger fragmented  lattices requires a large number of orbitals which in turn requires a lot of computational resources beyond our current capabilities. We believe, though, that these toy models captures the essence of the behavior of much larger systems.

\begin{figure}[] \setlength{\unitlength}{0.1cm}

\begin{picture}(75 , 25 ) 
{
\put(-3, 0)  {\includegraphics[ scale = 0.3]{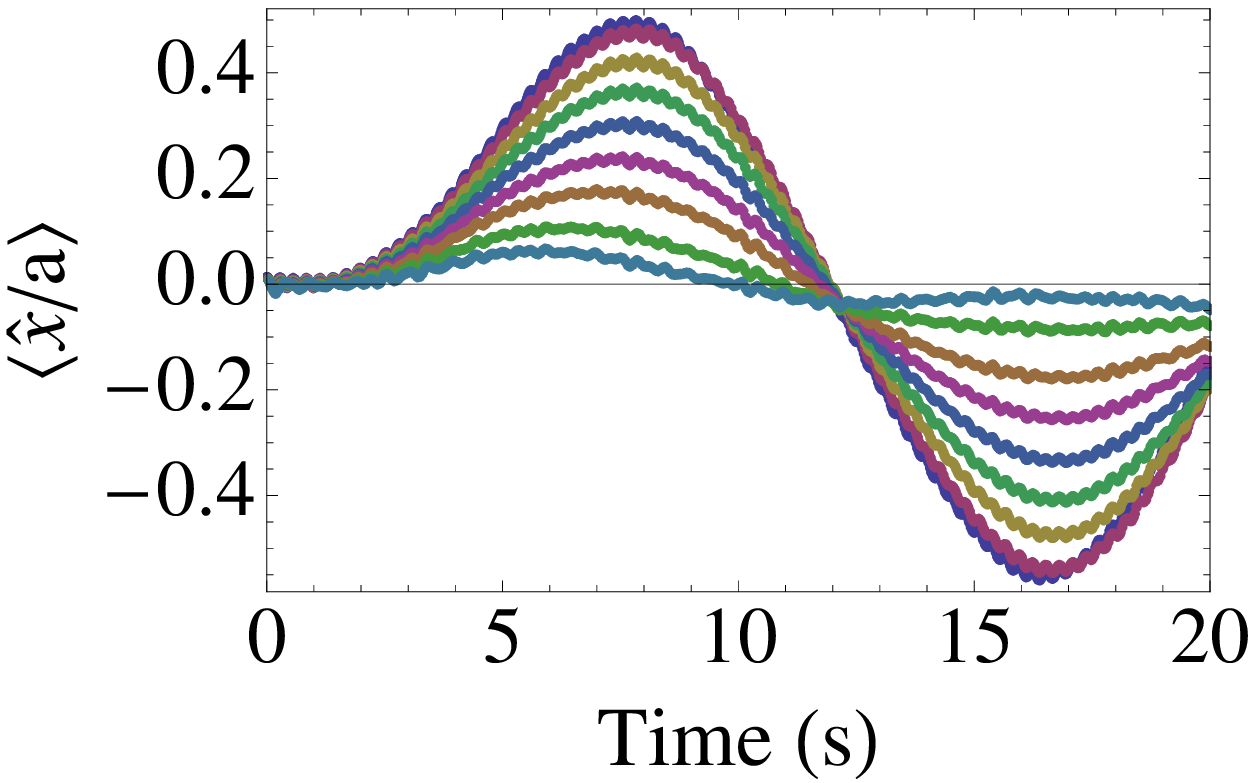}}
\put(39, 0)  {\includegraphics[ scale = 0.3]{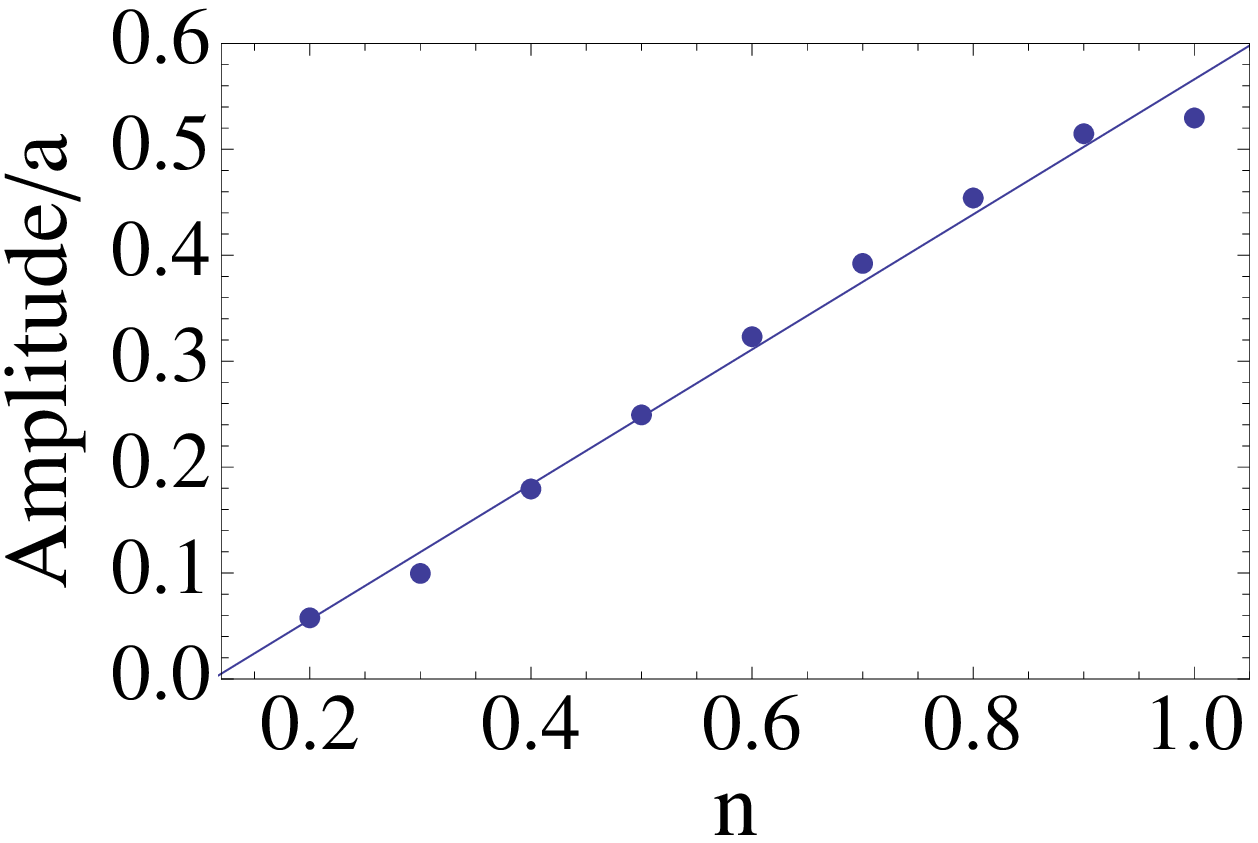}}

\scriptsize
\put(-3, 22.5) {(a)}
\put(37, 22.5) {(b)}
\normalsize
}
\end{picture} 

\caption{ \label{fig:Fig33}  Same as in Fig.  \ref{fig:Fig3} with the interatomic interaction $W(x_i-x_j)=0.1\times\delta(x_i-x_j)$. The  initial states are the same as in Fig.  \ref{fig:Fig3}.  (a) The oscillation of the atoms for different condensate fractions. (b) The  amplitude of the oscillations in (a) versus the condensate fraction $n$.}
\end{figure}

We repeated the previous simulation for a system with  interatomic interaction, $W(x_i-x_j)=0.1\times\delta(x_i-x_j)$. This value of $W_0$ is comparable to the interaction strength for 10 $^{87}$Rb atoms confined in a one-dimensional 10-site lattice \cite{beinke2017}.  The results presented in Fig. \ref{fig:Fig33} indicate that the small interaction has diminished the amplitude of the oscillations considerably, but the amplitude still exhibits a  linear dependence on the condensate fraction. For the sake of comparison, to generate an initial state with 90\% condensation, we used $W_0=0.55$.  We finally note that adding a realistic external trap potential to the  lattice potential should not change qualitatively the essential features distinguishing the superfluid and the Mott-like states.

\section{Quantum metrology using matter wave oscillation in optical lattices}\label{Sec4}

In this section, we explore the potential application of the atomic oscillation in an optical lattice orchestrated by the imprinted phase gradient  in the field of metrology.  Let us take the result derived in the previous section for the sinusoidal dependence of the current on the value of the imprinted phase step $\Delta \phi$ one step further and consider the interesting limit of a very large lattice (i.e., consisting of many wells) in a superfluid state. If a phase gradient $e^{ikx}$ is imprinted on the initial state of the condensate, the frequency of the matter wave density oscillation in the lattice will be a sinusoidal function of $ka$  where $a$ is the lattice constant. 

In order to prove this statement, consider first an optical lattice with periodic boundary conditions. The translation symmetry leads to the eigenstates forming a Bloch band structure. For a sufficiently deep optical lattice, where only overlap between Wannier states in neighboring sites is considered, the energy eigenvalues as a function of the Bloch quasi-momentum $q$ in the first Bloch band can be approximated by the tight binding model,  $E(q)=\alpha-2J\cos(qa) $ where $J$ is the tunneling coefficient and $\alpha$ is the on-site energy. In the first Brillouin zone, the  Bloch wave vector $q$  ranges from $-\pi/a$ to $\pi/a$ in discrete steps. 

For the more realistic case of rigid boundary conditions, or a lattice in an external trap,  the degeneracy between $q$ and $-q$ is removed and they are mixed into a new pair of eigenstates with different parity. If we keep $q$ as a label for these new states, the spectrum of the first band though remains more or less well approximated by the same formula for the tight-binding model, $E(q)=\alpha-2J\cos(qa)$.

Now, what happens when we initialize this system with a (boosted) state with momentum $k$, as we do by phase imprinting or by tilting the lattice? This state can be represented as a superposition of only a few eigenstates having the closest $|q|$ to $k$ by Fourier Sine/Cosine Series. Therefore, only those states with $|q|$ very close to $k$ will participate in the dynamics. For small momentum kicks that keep the dynamics within the first Bloch band, the timescale of  $\langle x\rangle$ can be roughly  determined by the differences of the eigenenergies of the closest eigenstates of different  parity  participating in the dynamics. 

For a periodic lattice, the difference between adjacent nondegenrate eigenstates is given by $\Delta_q \frac{\partial E}{\partial q}$, where $ \Delta_q=\frac{2\pi}{(N-1)a}$ and $N$ is the number of sites.
On the other hand and in the limit of a large lattice, the splitting between adjacent eigenstates for a nonperiodic lattice is approximately half the difference between adjacent nondegenrate eigenstates in the periodic case.
 Therefore,  the frequency $\omega$ of the oscillations of $\langle x\rangle$ can be approximated as
\begin{equation}
\hbar \omega \approx \frac{1}{2} \Delta_q 2Ja\sin(ka) =\frac{2\pi}{(N-1)}J\sin(ka)
 \label{omega}
\end{equation}
 The  maximum frequency occurs as expected around $ka = \pi/2$. For a double-well, in comparison, there are only two levels in each band, and hence the frequency of atomic oscillation is less sensitive to the value of $ka$.

It is noteworthy to highlight the difference between these oscillations which ensue in a periodic potential after a brief kick and the oscillations of quantum particles in periodic potential subject to a constant force, known as Bloch oscillations. Bloch oscillations are intrinsic and local oscillations; their frequency is defined by the lattice spacing and the force field while the oscillations considered here are global oscillations that sweep the whole lattice and therefore their frequency depends on the dimensions of the whole lattice. 

We analyzed the atomic oscillation in a superfluid optical lattice consisting of 32 sites of the form $V(x)=25\cos^2(\pi x)$ and initialized with different values of phase steps of $ka$. Since the energy eigenvalues in the first band fall in the range $[\alpha-2J,\alpha+2J]$, the tunneling coefficient $J$ can be estimated as $0.25\times(E_{31}-E_0)$ where $E_n$ is the $n$\textsuperscript{th} eigenenegry of the periodic lattice and equals $h\times 0.05\ \text{Hz}$.

\begin{figure}[] \setlength{\unitlength}{0.1cm}

\begin{picture}(75 , 25 ) 
{
\put(-4, 0)  {\includegraphics[ scale = 0.3]{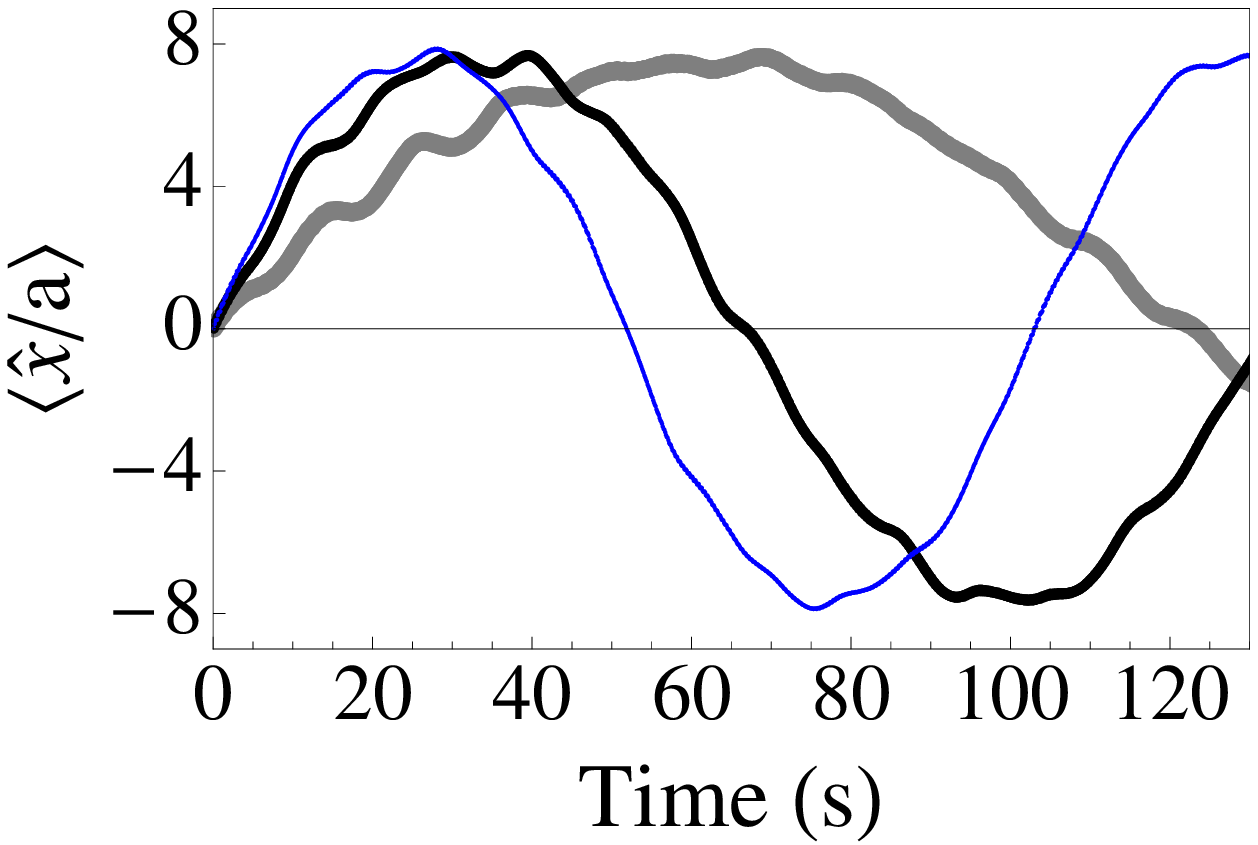}}
\put(39, 0)  {\includegraphics[ scale = 0.3]{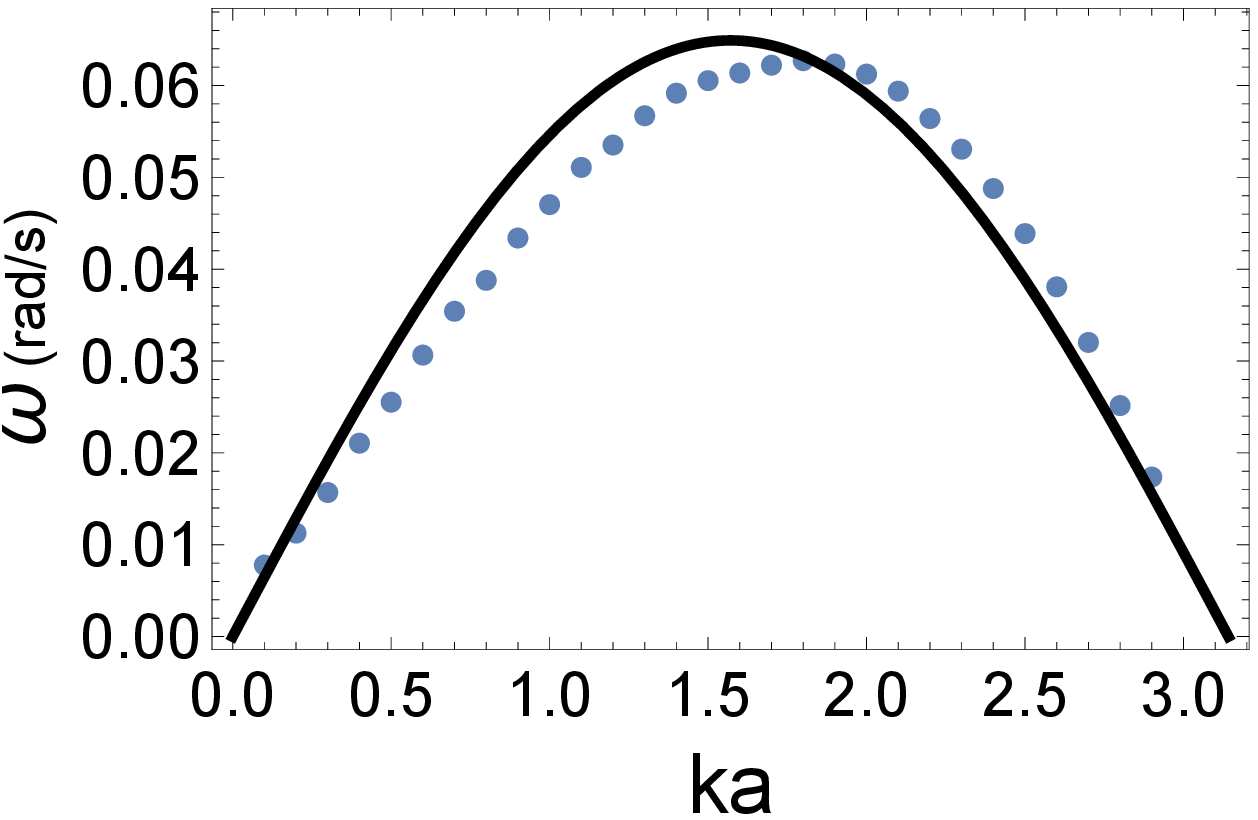}}

\scriptsize
\put(-3, 22) {(a)}
\put(37, 22) {(b)}
\normalsize
}
\end{picture} 

\caption{ \label{fig:Fig4} (Color online) (a) Atomic oscillation in a superfluid optical lattice consisting of 32 sites and initialized with  $ka= \Delta \phi=0.5$ (thick gray), 1.0 (black) and  1.5 (thin blue). (b) The dependence of the frequency of oscillation on the value of $ka$ in steps of of $ka=0.1$ obtained by numerical simulation using MCTDHB(1) (dotted) and compared with the theoretical prediction in Eq.~\ref{omega} (solid). }
\end{figure}

 In Fig.~\ref{fig:Fig4}-a, we show the matter wave oscillations for  $ka= \Delta \phi=0.5,\ 1.0, \ 1.5$. The dependence of the frequency on $ka$ for the range $0$ to $\pi$ is shown in Fig.~\ref{fig:Fig4}-b compared with the theoretical prediction in Eq.~\ref{omega}. We notice from the wiggling on top of the atomic oscillation in Fig. \ref{fig:Fig4}-a that higher band effects appear in the simulation.

The strong dependence of the oscillation frequency on the phase step $ka$ and the clear peaking around $ka=\pi/2$  suggests that such  a behavior can be used in quantum metrology, specifically as a magnetic or gravitational field gradiometer. Suppose that a field  in the $z$ direction  couples to the atoms and adds a potential term  $V=q \Gamma z$ to the Hamiltonian where $q$ is the charge (e.g., mass for gravitational field or magnetic dipole moment for a magnetic field) and $\Gamma$ is the field gradient in the $z$ direction. Let the 1D optical lattice be aligned in a direction orthogonal to the $z$ direction, say in the $x$ direction. If the lattice is physically tilted along the $z$ direction at an angle $\theta$ for a time interval $T$, each lattice site will acquire a phase at the end of the tilting interval according to the local strength of the field at that site while being tilted. The phase difference between neighboring sites is $\Delta \phi = q \Gamma a \sin(\theta) T/\hbar$. 

Consider first the case for the gravitational  field gradient where $\Gamma$ is the gravitational acceleration $g$.  This system is, somehow, the quantum  analog of the classical pendulum where the oscillation frequency is set by the gravitational field strength. Can we measure the gravitational acceleration $g$ by measuring the tilt angle that yields the maximum frequency of oscillation, corresponding to  $\Delta \phi =\pi/2$?
 Let us assume that Rubidium-87 atoms are trapped in  an optical lattice whose  wavelength is $780$ nm, leading to a lattice constant $a=\lambda/2=390$ nm. Since $\Gamma=g\approx9.8$ m/s\textsuperscript{2} and $q=$ mass(\textsuperscript{87}Rb), we find that $\sin(\theta) T$ should be of the order of $3\times 10^{-4}$ s. for $\Delta \phi =\pi/2$. The uncertainties in measuring $T$, $\theta$ and the frequency of oscillation, makes this method very inferior to the current methods of measuring $g$ using cold atoms  based on Bloch oscillations \cite{ferrari2006,poli2011} or based on atomic interference effects which attain resolutions of the order of $10^{-8}$ m/s\textsuperscript{2} \cite{abend2016,min2015}.

Next, let us consider magnetic field gradient measurement using Chromium atoms (\textsuperscript{52}Cr) which possess magnetic moment of 6 Bohr magneton trapped in an optical lattice whose wavelength $\lambda=$ 1064 nm \cite{griesmaier2005}. For a field gradient of 3000 nT/m, a value that can be encountered in the field of mineral exploration, we find that the naive estimation of $\sin(\theta) T$ should be of the order of 2  s. We notice that this value is much higher than the case for the gravitational potential measurement, making the accuracy for magnetic field gradiometry  much better than that for gravitational field gradiometry since the  effect of the uncertainty in measuring $\theta$ and $T$ will be smaller.  

To get a feeling of the actual values of the tilt angle $\theta$ and the tilt duration $T$, assume a tiny tunneling coefficient $ J=h\times 2.5$ Hz. The period of atomic oscillation at the maximum frequency in a 1000 site optical lattice can be  estimated from Eq. \ref{omega} to be around 64 s. Since the tilt interval should be much smaller than the period of oscillation, say at least by an order of magnitude, we obtain for  $\sin(\theta) T=2$ s and $T=4$ s   a tilt angle of 30 degrees. In order to exclude gravitational phase shift, this method is suitable only for horizontal field gradients, i.e., the tilt is made in a horizontal plane.

\section{Conclusion}\label{Sec5}

In this work we have proposed a  method for measuring the condensate fraction in optical lattices and multi-well traps. The method requires only giving the optical lattice a well-defined momentum boost or imprinting a well-defined phase gradient to different lattice sites, with the subsequent measurement of single-particle density in addition to identifying the behavior of the two extreme cases of the superfluid and the Mott insulator state. The prospects of utilizing this technique applied to superfluid states in the field of quantum metrology has been discussed.
Although we applied the technique to one-dimensional lattices only, it is generalizable to 2D and 3D lattices where the tilt mechanism and the subsequent transport of atoms can be independently performed in each direction.

\subsection*{Acknowledgments} 
T. A. Elsayed  thanks Prof. L. S.  Cederbaum for the discussion and for the hospitality of the ``Theoretical Chemistry Group" of Heidelberg University where part of this research was conducted.  The authors thank anonymous referees for their very helpful comments.  A. I. Streltsov cordially acknowledges the financial support by DFG. T. A. Elsayed acknowledges the financial support by Villum Foundation.

\bibliographystyle{naturemag}

\bibliography{cold}

\end{document}